\documentclass[a4paper,11pt]{article}
\usepackage{color,graphicx}
\usepackage{amsmath}
\title{  Non-resonant four-body decay of $B \to D^- \pi^+\pi^+\pi^- $ }
\author{Mohammad Rahim Talebtash\footnote{mtalebtash@students.semnan.ac.ir},
 Hossein Mehraban\footnote{hmehraban@semnan.ac.ir}\\
Physics Department, Semnan University\\
P.O.Box 35195-363, Semnan, Iran}
\begin{document}
\maketitle
\begin{abstract}
We calculate the branching ratio of the non-resonant $B \to D^- \pi^+\pi^+\pi^- $ decay using a simple model based on the framework of the factorization approach where one can ommit the nonfactorization effects. In naive factorization approach, there are only two tree diagrams for this decay mode. In the first diagram, the matrix element of decay mode is factorized into a $B\to D$ form factor multiplied by a $3\pi$ decay constant and in the second diagram, the matrix element is factorized into a $B\to D\pi$ form factor multiplied by a $2\pi$ decay constant, We assume that in the rest frame of B meson,  the $D$ meson remains stationary, so we obtain the value $(3.47\pm0.14)\times 10^{-3}$ for the branching ratio of the $B \to D^- \pi^+\pi^+\pi^- $ decay mode, while the experimental results are $(3.9\pm1.9)\times10^{-3}$.
\end{abstract}
\section{Introduction}
The three-body and four-body decays of $B$ meson are more complicated than two-body case, specifically where both non-resonant and resonant contribution exist. In this types  the theoretical momentums of the output particles are not directly calculable. In four-body cases, the momentums and form factors are written in terms of five variables: two invariant-mass of particles variable and three angles.\\
In the previous works ~\cite{1,2}, for some specific three-body decays assumed that one of the output mesons due to its heavy weigh, remains stationary and the two other output mesons move back to back and the amplitude of this three decay modes can be calculate. In this work we extended this assumption to non-resonant four-body decays and calculate the branching ratio of  $B \to D^- \pi^+\pi^+\pi^- $ decay in naive  factorization approach where we can ignore the non-factorization effects. We assume that the $D$ output meson, due to its heavy weight, remains stationary and the three pion mesons move in a plane and we have directly resolved their momentums and we calculated the branching ratio of $D^- \pi^+\pi^+\pi^-$ and predicted $(3.47\pm0.14)\times 10^{-3}$ which is compatible with the experimental result $(3.9\pm1.9)\times 10^{-3}$.\\
This paper is organized as follows. In section 2 we consider the amplitudes of the  $D^- \pi^+\pi^+\pi^-$  decay and we present the transition matrix needed  and  the kinematics of the four-body decays.  In section 3 we give the numerical results, and in the last section we
have a summary.
\section{Non-resonant amplitudes of the $B \to D^-\pi^+\pi^+\pi^-$}
For the four-body $B \to D^-\pi^+\pi^+\pi^-$, the Feynman diagrams are shown in Fig. 1, where both panel show the $\bar{b}\to \bar{c}u\bar{d}$ tree transitions~\cite{3,4,5}. Under the factorization approach, the contribution of the diagram (a) in decay amplitude consists of the  $<B\to D^->\times<0\to\pi\pi\pi>$ distinct factorization term, where $<B\to D>$ denotes the form factor of $B$ to $D$  meson transitional, and the contribution of the diagram (b) consists of a $<B\to \pi\pi>\times <0\to \pi \pi>$ distinct factorization term, where $<B\to \pi\pi>$ denotes the two-meson transition matrix element.
\begin{figure}[h]
\centering
\includegraphics [scale=0.4]{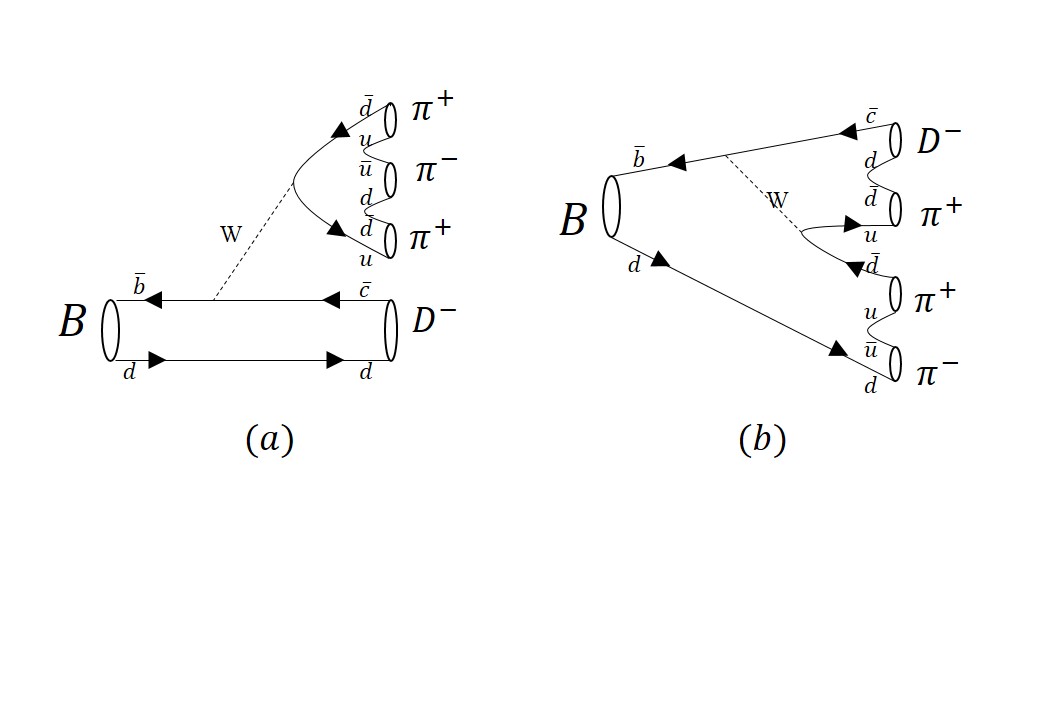}
 \caption{Feynman diagram for the $B \to D^- \pi^+\pi^+\pi^- $ decay.}
\label{fig1}
\end{figure}
 So the matrix elements of the $B \to D^-\pi^+\pi^+\pi^-$ decay amplitudes are given by
\begin{eqnarray}
|\mathcal{M}|&=& i \frac{G_F}{\sqrt{2}} V_{cb}V_{ud}\Big{\{}a_1<\pi\pi\pi|(\bar{d}u)_-|0><D|(\bar{c}b)_-|B>\\ \nonumber
&&+a_2<D\pi|(\bar{c}b)_-|0><\pi\pi|(\bar{d}u)_-|B>\Big{\}},
\end{eqnarray}
Where in factorization, the general expression of the $a_i$ quantities in terms of effective Wilson coefficients is~\cite{ai}
\begin{eqnarray}
a_1&=&c_1+\frac{1}{3}c_2,\nonumber \\
a_2&=&c_2+\frac{1}{3}c_1.
\end{eqnarray}
\subsection{Form factors and decay constants}
According to  Fig. 1(a), the form factor of $B$ meson  to $D$ meson is defined as fallow~\cite{6}
\begin{eqnarray}
<D(p_4)|(\bar{c}b)_{(V-A)}|B(P)>&=&\Big{[}(p_B+p_4)_\mu-\frac{m_B^2-m_D^2}{(p_B-p_4)^2}(p_B-p_4)_\mu\Big{]}F_1(q^2) \nonumber \\
&&+\frac{m_B^2-m_D^2}{(p_B-p_4)^2}(p_B-p_4)_\mu F_0(q^2),
\end{eqnarray}
and the point like 3-body matrix element is assumed to be ~\cite{7}

\begin{eqnarray}
<\pi(p_1)\pi(p_2)\pi(p_3)|(\bar{d}u)_{(V-A)}|0>&=&\frac{2i}{f_\pi}\Big{[}p_{2\mu}-\frac{(p_B-p_4).p_2}{(p_B-p_4)^2-m_\pi^2}(p_B-p_4)_\mu\Big{]}. \nonumber \\
\end{eqnarray}
According to Fig. 1(b), for the current-induced process, the two-meson transition matrix element $<\pi\pi|(\bar{d}u)_{(V-A)}|B>$ has the fallowing general expression ~\cite{7,8}

\begin{eqnarray}
<\pi\pi|(\bar{d}u)_{(V-A)}|B>&=& i r (p_B-p_1-p_2)_\mu+i\omega_+(p_2+p_1)_\mu+i\omega_-(p_2-p_1)_\mu, \nonumber \\
\end{eqnarray}
where the form factors $r, \omega_+$ and $\omega_-$ for the non-resonant decay are given by
\begin{eqnarray}
r &=& \frac{f_B}{2f_\pi^2}-\frac{f_B}{f_\pi^2}.\frac{p_B.(p_2-p_1)}{(p_B-p_1-p_2)^2-m_B^2}+\frac{2g f_{B_S^*}}{f_\pi^2}.\sqrt{\frac{m_B}{m_{B_S^*}}}.\frac{(p_B-p_1).p_1}{(p_B-p_1)^2-m_{B_s^*}^2} \nonumber \\
&&+\frac{4g^2f_B}{f_\pi^2}.\frac{m_Bm_{B_S^*}}{(p_B-p_1-p_2)^2-m_B^2}.\frac{p_1.p_2-p_1.(p_B-p_1)p_2.(p_B-p_1)/m_{B_s^*}^2}{(p_B-P_1)^2-m_{B_S^*}^2}, \nonumber \\
\end{eqnarray}
\begin{eqnarray}
\omega_+&=&-\frac{g}{f_\pi^2}.\frac{f_{B_S^*}m_{B_S^*}\sqrt{m_Bm_{B_S^*}}}{(p_B-p_1)^2-m_{B_S^*}}\Bigg{[}1-\frac{(p_B-p_1).p_1}{m_{B_S^*}^2}\Bigg{]}+\frac{f_B}{2f_\pi^2},
\end{eqnarray}
\begin{eqnarray}
\omega_-&=&-\frac{g}{f_\pi^2}.\frac{f_{B_S^*}m_{B_S^*}\sqrt{m_Bm_{B_S^*}}}{(p_B-p_1)^2-m_{B_S^*}}\Bigg{[}1+\frac{(p_B-p_1).p_1}{m_{B_S^*}^2}\Bigg{]}.
\end{eqnarray}
The tow-body $<D\pi|(\bar{c}b)_{(V-A)}|0>$ matrix element can be related to the $D$ and $\pi$ matrix element of the weak interaction current:
\begin{eqnarray}
<D(p_4)\pi(p_3)|(\bar{c}b)_{(V-A)}|0>&=&(p_4-p_3)F^{D\pi}(q^2),
\end{eqnarray}
where $F^{D\pi}(q^2)$ is the $D$ to $\pi$ transition form factor and determined from experiment~\cite{9}.
\subsection{The kinematics of four-body decay}
For four-body  $B$ decays, we can work  in the rest frame of the B meson. By following the definition given in ~\cite{10,11}, when B meson
decays to final-state mesons, the kinematics of the four-body decays can be described in term of five independent variables where the  momenta and masses are defined as
\begin{eqnarray}
P(p,M)&\to&P_1(p_1,m_1)P_2(p_2,m_2)P_3(p_3,m_3)P_4(p_4,m_4) ,
\end{eqnarray}
with $p^2=M^2$ and $p_i^2=m_i^2$ where $i=1,2,3$ and $4$. We choose the set of independent variables as $\{ s_{12},s_{34},\theta_1,\theta_3,\phi\}$ which have the following geometrical meaning (see Fig. (2))
\begin{itemize}
\item$s_{12} = (p_1+ p_2)^2$, is the invariant-mass of particles 1 and 2;
\item$s_{34} = (p_3 + p_4)^2$, is the invariant-mass of particles 3 and 4;
\item$\theta_1 (\theta_3)$, is the angle between the three-momentum of particle 1 (particle 3) with respect
to the direction of $\vec{p}_{12} =\vec p_1 + \vec p_2$ ($ \vec p_{34} = \vec p_3 +\vec p_4)$ defined in the rest frame
of the decaying particle (see fig.(2));
\item$\phi$ is the angle between the planes defined by particles (1, 2) and (3, 4) also in the rest
frame of the decaying particle.
\end{itemize}
\begin{figure}[h]
\centering
\includegraphics [scale=0.4]{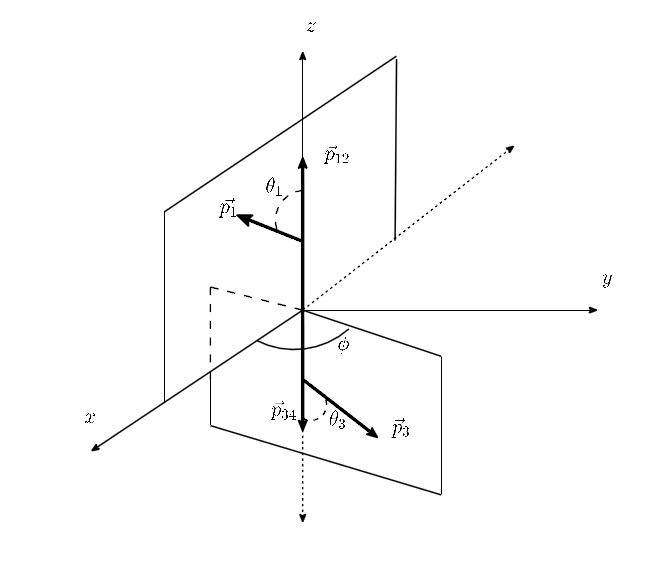}
 \caption{Kinematics of four-body decays in the rest frame of the decaying particle, $\Sigma_{i=1}^4\vec{P}_i=0$.
We have defined $\vec{p_{ij}} = \vec {p_i} + \vec{p_j}$, such that $\vec{p_{12}} + \vec{p_{34}} = 0$.}
\label{fig2}
\end{figure}
With this choice of kinematics, the differential decay rate in the rest frame of the decaying
particle can be written as:
\begin{eqnarray}
d\Gamma&=&\frac{X\beta_{12}\beta{34}}{4(4\pi)^6m_B^3}\bar{|\mathcal{M}|^2}\frac{1}{n!}ds_{12}ds_{34}d\cos \theta_1d\cos \theta_3d\phi,
\end{eqnarray}
where $\beta_{12}(\beta_{34}$) is the velocity of particle 1 (particle 3) in the center of mass frame of particles
1 and 2 (3 and 4) and $X=[(P^2-S_{12}-S_{34})^2-4S_{12}S_{34}]^{1/2}$, the $n!$ factor
accounts for identical particles in the phase space and $\bar{|\mathcal{M}|}$ is the amplitude of the decay  (Eq.(1)).\\
For considering the $B \to D^- \pi^+\pi^+\pi^- $ decay mode, we work in the rest frame of the $B$ meson and assume that the $D^-$ meson due to its heavy weight, has a small momentum and can easily be neglected, so the three pions move in a plane. Hence the  momentums of  $B$ meson and final-state mesons are given by
\begin{equation*}
\begin{array}{cc}
p_{1}=(p_{\pi1}^0,\vec{p}_{\pi1}),& p_4=(m_D,\vec{0}),\\
p_{2}=(p_{\pi2}^0,\vec{p}_{\pi2}),  &p_B=(m_B,\vec{0}) \\
p_{3}=(p_{\pi3}^0,\vec{p}_{\pi3}),
\end{array}
\end{equation*}
where $\vec{p}_{\pi1}=\vec{p}_{\pi2}=\vec{p}_{\pi3}$ and $p_{\pi1}^0=p_{\pi2}^0=p_{\pi3}^0=(m_B-m_D)/3$.
Now we can easily integrate the decay rate with respect to all independent variables and we obtain a simple experiment
\begin{eqnarray}
\Gamma&=&\frac{X\beta_{12}\beta_{34}}{4(4\pi)^6m_B^3}\bar{|\mathcal{M}|^2}\frac{8\pi}{3!}s_{12}s_{34}.
\end{eqnarray}
\section{Numerical results}
The Wilson coefficients $c_i$ have been calculated in different schemes. In this paper we use consistently the naive dimensional regularization (NDR) scheme. The values of $c_i$ at the scales $\mu=m_b$ at the leading order (LO) are ~\cite{12}

\begin{equation*}
\begin{array}{cc}
c_1=1.117 & c_2=-0.268 .
\end{array}
\end{equation*}
The parameter $g$ in the form factors,
determined from the $D^*\to D \pi$ decay is
\begin{eqnarray}
g&=&0.3\pm0.1.
\end{eqnarray}
For the elements of the CKM matrix, we use the values of
the Wolfenstein parameters and obtain~\cite{13}
\begin{equation*}
\begin{array}{cc}
  V_{ud}=0.9745 & V_{cb}=0.0415
\end{array}
\end{equation*}
The meson masses, decay constants and form factors needed in our
calculations are taken as (in units of GeV)~\cite{14,15}
\begin{equation*}
\begin{array}{cccc}
m_B=5.279 &m_{B^*_s}=5.413 & m_D=1.87 & m_\pi=0.139 \\
f_B=0.176 & f_{B^*_s}=0.220 & f_\pi=0.13 & F^{B\to D}_0=0.52 \\
F^{B\to D}_1=0.52.
\end{array}
\end{equation*}
Using the parameters relevant for the  $B \to D^- \pi^+\pi^+\pi^- $ decay, we calculate the branching ratio of this decay which is
\begin{eqnarray}
B(B \to D^- \pi^+\pi^+\pi^-)=(3.47\pm0.14)\times 10^{-3}
\end{eqnarray}
where the experimental value is~\cite{14}
 \begin{eqnarray}
B(B \to D^- \pi^+\pi^+\pi^-)_{expt}=(3.9\pm1.9)\times 10^{-3}.
\end{eqnarray}
\section{Summary}
In this work, we have considered four-body
$B\to D^- \pi^+\pi^+\pi^-$  decays under the assumption that the
$D$ meson has small and negligible momentum due to
its heavy weight. This decay mode have contribution of $a_1$ and $a_2$ coefficients. According to the Feynman diagrams, each contribution factorized into a form factor and a decay constant. Under the assumption of negligible momentum of D meson, the three output pion mesons move in a plane and we have directly resolved their momentums and we calculated the branching ratio of $D^- \pi^+\pi^+\pi^-$ and predicted $(3.47\pm0.14)\times 10^{-3}$ which is compatible with the experimental result $(3.9\pm1.9)\times 10^{-3}$.

\end{document}